\documentclass[preprint,12pt]{elsarticle}

\usepackage{amssymb}
\usepackage{siunitx}
\usepackage{hyperref}
\usepackage{xcolor}

\journal{Journal of Quantitative Spectroscopy and Radiative Transfer}

\begin{document}

\begin{frontmatter}



\title{Extension of Radiative Transfer Coherent Backscattering RT-CB code to dense discrete random media}


\author[inst1,inst2]{Johannes Markkanen}

\affiliation[inst1]{organization={Institute of Geophysics and Extraterrestrial Physics, TU Braunschweig, Germany},
            addressline={Mendelssohnstrasse 3}, 
            city={Braunschweig},
            postcode={38106}, 
            country={Germany}}
\affiliation[inst2]{organization={Max Planck Institute for Solar System Research, Goettingen, Germany},
            addressline={Justus-von-Liebig-Weg 3,}, 
            city={Goettingen},
            postcode={37077},
            country={Germany}}

\author[inst3]{Antti Penttil\"a}

\affiliation[inst3]{organization={Department of Physics},
            addressline={PO box 64},
            city={University of Helsinki},
            postcode={00014},
            country={Finland}}

\begin{abstract}
The Radiative transfer coherent backscattering (RT-CB) code is extended to apply to dense discrete random media of optically soft spherical particles. This is achieved by utilizing the well-known static-structure-factor (SSF) correction derived from the Percus-Yevick approximation for sticky-hard-sphere (SHS) pair correlation function. The code is verified against the numerically exact electromagnetic method for small spherical media consisting of submicrometer-sized icy particles at optical wavelengths. The SSF-corrected RT-CB method significantly improves  solution accuracy, and the solution agrees well with the numerically exact solution when the packing density is less than 20\% and particles are optically soft, i.e., the refractive index of particles is close to that of the background medium. 
\end{abstract}




\end{frontmatter}


\section{Introduction}

Light scattering by macroscopic objects consisting of densely-packed microscopic particles is an open computational problem.  Numerically exact electromagnetic methods solving Maxwell's equations such as the discrete-dipole approximation, volume-integral-equation, finite-element, finite-difference, and superposition T-matrix methods can solve problems up to a few tens of wavelengths in size and containing less than a few million small particles \cite{penttila2021}. For larger problems, approximate methods are needed.

Numerical methods based on the radiative transfer equation (RTE) are powerful for analysing scattering by clouds of particles in which the distances between the particles are large. The RTE for sparse medium has been derived from Maxwell's equations under the ladder approximation and is theoretically well understood \cite{MISHCHENKO2016, DOICU2019}. Further, the coherent backscattering effect has been incorporated into radiative transfer algorithms by accounting for cyclical diagrams that are neglected in the derivation of the standard RTE \cite{TISHKOVETS2003, Muinonen2004, TISHKOVETS2020}. The Monte Carlo algorithm, known as radiative transfer coherent backscattering (RT-CB) \cite{Muinonen2004}, is applicable to a finite medium of spherical particles allowing for a direct comparison between the results of RT-CB and numerically exact methods. Interestingly, even though RT-CB assumes that the medium is large, the solutions for small ($\sim$10 wavelengths) sparsely-packed spherical media agree well with the solutions of the exact numerical methods \cite{Muinonen2012, vaisanen2016}. This allows us to study the validity regime of the RT-CB solution. Unfortunately, RTE-based methods are not applicable to dense media due to various assumptions made in the derivation of the RTE \cite{MISHCHENKO2016}.

Several attempts have been made to extend the applicability of the RTE to dense discrete random media \cite{Cartigny1986,Drolen1987, Tsang1985, MISHCHENKO1994, Tsang2000, TISHKOVETS2006, Tsang2007}.  Most notably, accounting for the correlated positions of particles in terms of the Percus-Yevick approximation for the pair correlation function of hard spheres has been widely applied in remote sensing applications. Such an approach gives rise to various forms of the dense media radiative transfer (DMRT) equation \cite{Tsang1985, Tsang2000, Tsang2007}, or the static-structure-factor (SSF) correction \cite{Cartigny1986, MISHCHENKO1994, ito2018, Ma2020}. The Percus-Yevick approximation provides an analytical solution for the pair-distribution function resulting in a fast numerical solution. However, these approaches do not account for the coherent backscattering effect, and the results have never been compared to those of numerically exact methods. 
Another approach is to numerically compute the ensemble-averaged Mueller matrix, scattering and extinction coefficients of an elementary volume containing a sufficient number of particles to account for the near-field effects. The ensemble-averaged scattering properties of the elementary volume are then used as inputs in the standard radiative transfer solution \cite{Zurk1996, TISHKOVETS2013, ito2018, Markkanen2018_2}. Also, a numerical solution in terms of incoherent volume elements using T-matrices has been proposed \cite{Muinonen2018, Markkanen2018, Vaisanen2019}. This method, unlike the other above-mentioned methods, does not use the far-field approximation in the computations of interactions between the elementary volumes in the scattering sequences. Instead, it uses a numerically exact approach based on the T-matrix representation of scatterers and the translation addition properties of spherical vector wave functions. The method, however, is computationally expensive as it requires solving a large number of T-matrices for the elementary volumes and their interactions in the scattering sequences. 

In this paper, we apply the sticky-hard-sphere (SHS) Percus-Yevick pair correlation function \cite{Baxter1968} and incorporate it into the Monte-Carlo RT-CB code. The extended version of the RT-CB code is open source and available in \url{https://bitbucket.org/planetarysystemresearch/rtcb_public/}. We show that the SSF correction improves the accuracy of the standard RT-CB solution for a medium of densely-packed optically soft particles. Furthermore, the implemented SHS model allows us to treat aggregated particulate media where the adhesive forces between particles can be controlled by the stickiness parameter.

\section{Radiative transfer coherent backscattering RT-CB}

RT-CB is a numerical solution based on Monte Carlo integration of ladder and cyclical diagrams derived from the theory of radiative transfer and coherent backscattering \cite{Muinonen2004, Muinonen2012}. The RT-CB code traces rays characterized by the Stokes parameters within the medium. Each scattering process is described by the single scattering properties, i.e., the amplitude scattering matrix elements $S_1$ and $S_2$, and the single scattering albedo $\omega$. The scattered Stokes parameters after each scattering process are updated at the fixed angles, allowing for the computation of the exact phase differences between the forward and backward propagating scattering paths that give rise to the coherent backscattering effect. Exponential attenuation is assumed with the mean free path length $\ell$.

\subsection{Static structure factor correction}

The dense medium radiative transfer equation resembles that of the standard RTE except the input parameters are modified by the scattering angle ($\theta$) dependent static structure factor $S_{\rm M}(\theta)$ \cite{Cartigny1986,MISHCHENKO1994, Tsang2001}. The SSF correction accounts for the collective scattering and interference effects arising from the correlated positions of particles in a dense medium. The pair distribution function $g(r)$, where $r$ is the distance between the centers of a particle pair, describes correlation in the particle positions and it has a closed-form analytical expression under the Percus-Yevick approximation for hard spheres \cite{Wertheim} and sticky hard spheres \cite{Baxter1968}. The Fourier transform of the pair distribution function is related to the static structure factor.

The SSF-corrected Mueller matrix for spheres is given by
\begin{equation}
    M_{ij}^{\rm SSF}(\theta) = S_{\rm M}(\theta) M_{ij}^{\rm Mie}(\theta), \,\,\, i,j = 1,2,3,4  
\end{equation}
where $M_{ij}^{\rm Mie}$ is the Lorenz-Mie Mueller matrix. The Lorenz-Mie Mueller matrix is non-depolarizing, hence it can be described with a diagonal amplitude scattering matrix with elements $S_1$ and $S_2$\cite{Hovenier94}.
The RT-CB code uses the amplitude scattering matrix instead of the Mueller matrix in order to treat coherent backscattering. The SSF-corrected amplitude scattering matrix elements read as
\begin{equation}
 S_i^{\rm SSF}(\theta) = \sqrt{S_{\rm M}(\theta)} S_i(\theta), \,\,\, i = 1,2.     
\end{equation}

The SSF-corrected scattering cross section is given by
\begin{equation}
C^{\rm SSF}_{\rm sca} = \frac{C_{\rm sca}^{\rm Mie}}{4\pi}\int_{4\pi} S_{\rm M}(\theta) M^{\rm Mie}_{11}(\theta) \, d\Omega     
\end{equation}
and the extinction cross section by
\begin{equation}
    C_{\rm ext}^{\rm SSF} = C^{\rm SSF}_{\rm sca} + C^{\rm Mie}_{\rm abs} 
\end{equation}
where $C^{\rm Mie}_{\rm abs}$ is the absorption cross section from the Lorenz-Mie solution. Then, the SSF-corrected albedo and the mean free path read as
\begin{equation}
    \omega^{\rm SSF} = \frac{C^{\rm SSF}_{\rm sca}}{C^{\rm SSF}_{\rm ext}},
\end{equation}
\begin{equation}
    \ell^{\rm SSF} = \frac{4\pi a^3}{3 C_{\rm ext}^{\rm SSF}}f,
\end{equation}
where $a$ is the monomer radius and $f$ is the volume fraction.

We implemented the SSF correction to RT-CB by employing the Percus-Yevick approximation for the SHS model \cite{Baxter1968}. The SHS model has three parameters, the monomer size parameter $ka$, volume fraction $f$ and the stickiness parameter $\tau$. The SHS model is defined by the following square-well potential 
  \begin{equation}
    u(r) = \left \{
    \begin{array}{lll}
       \infty &\text{for}& 0 < r < s \\
       \ln\frac{12\tau(2a-s)}{2a} &\text{for}& s < r < 2a \\
       0 &\text{for}&  r > 2a
    \end{array}
    \right.
    \label{eq_potential}
  \end{equation}
  in the limit
  \begin{equation}
\lim_{s\rightarrow 2a} (2a - s) e^{-u(r)} = \frac{2a}{12\tau} < \infty.
  \end{equation}
The square-well potential function has a nature of surface adhesion in which the dimensionless parameter $\tau^{-1}$ is proportional to the adhesion strength. The main advantage of using such a simple potential function is that it allows for a closed form expression for the pair distribution function and the structure factor under the Percus-Yevick approximation \cite{Baxter1968}.

The closed-form analytic expression for the structure factor reads as \cite{Tsang2001} (Eqs. 8.4.19–8.4.22) 
\begin{equation}
S_M(f,\theta,\tau) = [A(X)^2 + B(X)^2]^{-1} 
\end{equation}
\begin{equation}
A(X) = \frac{f}{1-f} [(1-tf + \frac{3f}{1-f})\Sigma(X) + (3-t(1-f))\Psi(X)] + \cos(X)
\end{equation}
\begin{equation}
B(X) = \frac{f}{1-f}X\Sigma(X) + \sin(X)    
\end{equation}
\begin{equation}
    \Sigma(X) = 3[\frac{\sin(X)}{X^3} - \frac{\cos(X)}{X^2}]
\end{equation}
\begin{equation}
\Psi(X) = \frac{\sin(X)}{X}    
\end{equation}
\begin{equation}
    X = 2 ka \sin{(\frac{\theta}{2})} 
\end{equation}
in which the parameter $t$ is the smallest solution of the following equation
\begin{equation}
    \frac{f}{12}t^2 - (\tau+\frac{f}{1-f})t + \frac{1+f/2}{(1-f)^2} = 0.
\end{equation}
When the stickiness parameter $\tau\rightarrow\infty$, $t\rightarrow 0$ and the expression is reduced to the structure factor of the non-sticky hard sphere model.

\begin{figure*}[htb]
      \includegraphics[width=0.5\textwidth]{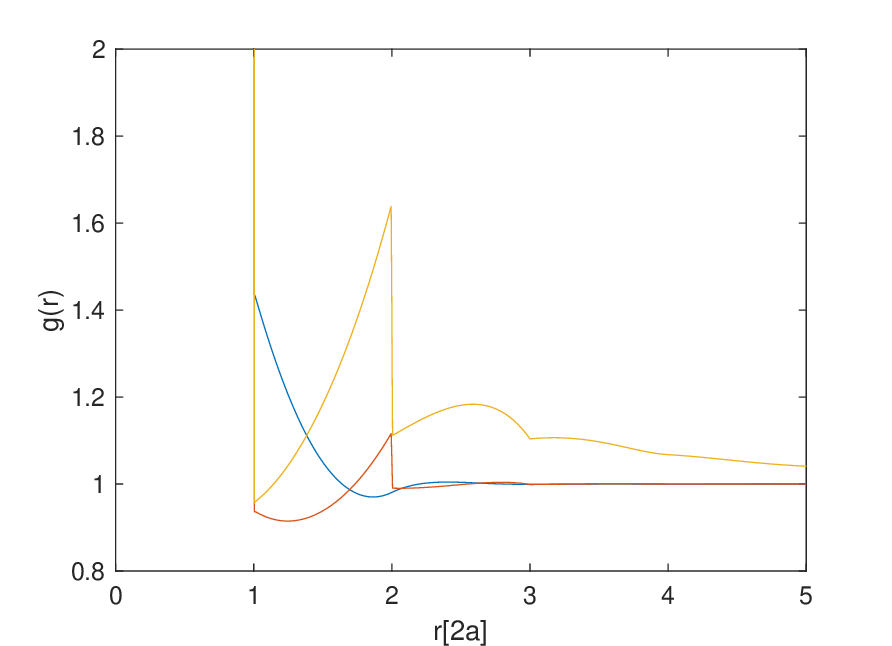}
      \includegraphics[width=0.5\textwidth]{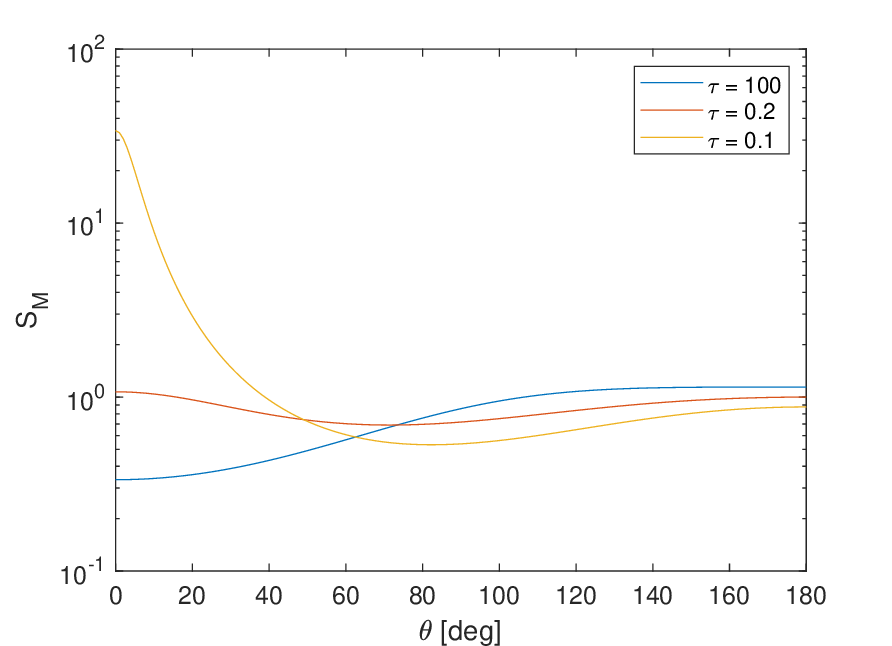}
      \caption{Pair distribution function (left) and the corresponding static structure factor (right) calculated using the SHS Percus-Yevick approximation for f=0.14 and ka = 1.5 with the three different stickiness parameters $\tau$.}
      \label{fig0}
\end{figure*}

Fig. \ref{fig0} shows an example of the pair distribution functions $g(r)$ and the corresponding static structure factors $S_{\rm M}(\theta)$ for $ka = 1.5$, $f = 0.14$ and $\tau = 100$, 0.2 and 0.1. The pair distribution function $g(r) = 0$ when $r<2a$ as the spheres are not allowed to overlap. For small $\tau$, $g(2a)$ reaches a high value indicating a high probability of finding a sphere in contact with the reference sphere. With increasing $\tau$, $g(2a)$ decreases as the attractive potential between the monomers becomes weaker.

  \subsection{Placement algorithm for sticky spheres}
  \label{sec_placement}
  To get more insight into the internal structure of the medium following the SHS model, we briefly review a particle placement algorithm proposed in \cite{Seaton1987, Kranendonk1988}. An implication of the potential defined in (\ref{eq_potential}) is that the probability of a test particle to form an $\alpha$-bond is proportional to $(2a/(12\tau))^\alpha$. Here, $\alpha = 0$ bond means that the test particle is not connected to any other particle, $\alpha = 1$ is connected to one another particle, $\alpha = 2$ is connected to two other particles, and so on.
A total effective volume related to $\alpha$-bond of the test particle is given by
  \begin{equation}
    \displaystyle
    V^\alpha_{eff} = \sum_i (\frac{2a}{12\tau})^\alpha \int dq_i
  \end{equation}
  where $dq_i$ is all the degrees of freedom available after making the bond, and the summation is over all the other particles that can make the $\alpha$-bond. Then, the probability of changing the state of the configuration to a new configuration is given by
  \begin{equation}
P^{\alpha\rightarrow\alpha'} = \frac{V^{\alpha'}_{eff}}{\sum_{\alpha} V_{eff}^\alpha}.
  \end{equation}
  The acceptance-rejection Monte Carlo placement algorithm in \cite{Kranendonk1988} operates as follows: First, fill a medium of radius $R$ with $N$ spheres such that the volume fraction f is reached. Second, select a test particle and compute the effective volumes  $V^\alpha_{eff}$ for the bonding states $\alpha = 0$ to 3. Third, move the test particle to a new state $\alpha'$ based on the probability $P^{\alpha\rightarrow\alpha'}$. Accept the move if the test particle does not overlap with any other spheres and stays inside the medium of radius $R$. Select another test particle and repeat.

The effective volume for the 0-bond
\begin{equation}
  V_{\rm eff}^0 = 4/3\pi R^3
 \end{equation}
since an unbounded test particle can be placed anywhere in the medium including overlapping regions. A single bond can be created anywhere where the distance between the centers of test particle and another particle in the system equals twice the sphere radius $a$. Hence, the effective volume reads as 
\begin{equation}
  V^{1}_{\rm eff} = (N-1) \frac{2a}{12\tau} 4\pi (2a)^2.
\end{equation}
The required condition to create a double bond is that the distance between a pair of particles $d \leq 4a$. The test particle can be placed in a circle where the distances between the center of the test particle and the center both of particles in the pair equal $2a$ resulting in
\begin{equation}
V^2_{\rm eff} = (\frac{2a}{12\tau})^2 \sum 2\pi\sqrt{(2a)^2-(\frac{d}{2})^2},
\end{equation}
where the summation is over all the particle pairs with $d \leq 4a$. The effective volume associated with the triple bond is given by
\begin{equation}
V^3_{\rm eff} = \sum (\frac{2a}{12\tau})^3 \frac{1}{|\hat{e}_{ik} \cdot (\hat{e}_{jk} \times \hat{e}_{lk})|}  
\end{equation}
in which $\hat{e}_{ik}$, $\hat{e}_{jk}$, and $\hat{e}_{lk}$ are the unit vectors joining the test particle center to the centers of other particles in a triplet. The summation is over all the triplets capable to create the triple bond with the test particle. Finally, if the test particle is already connected to 4 or more particles it is not allowed to break the bond.

The effective volumes $V^0_{\rm eff}$ and $V^1_{\rm eff}$ are independent on the current configuration of the system whereas $V^2_{\rm eff}$ and $V^3_{\rm eff}$ depend on the configuration. Therefore, the lists of possible pairs and triples that can form 2 and 3 bonds are updated after every successful move in the algorithm. The described algorithm requires thousands of trial moves per particle to reach the equilibrium state. Thus, it becomes computationally extremely slow for large number of particles with small $\tau$.  A more detailed description of the placement algorithm and its performance can be found in \cite{Seaton1987, Kranendonk1988, Tsang2001}. 

\begin{figure}[htb]
  \includegraphics[width=0.75\textwidth]{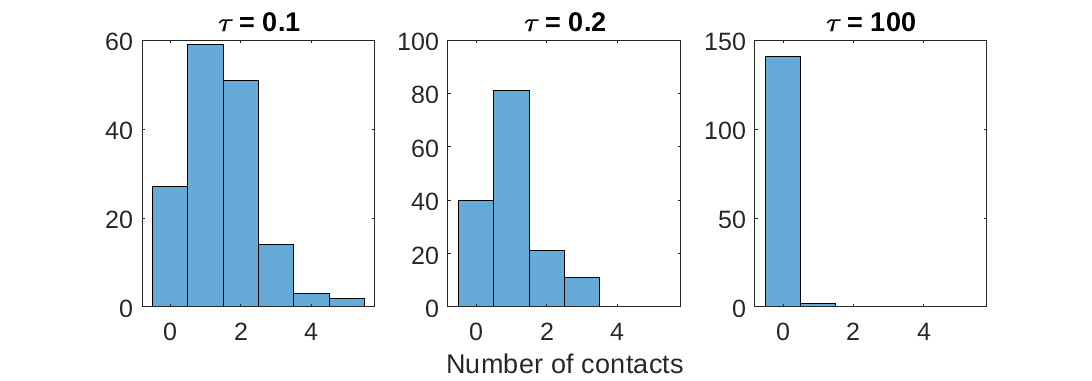}
   \includegraphics[width=0.75\textwidth]{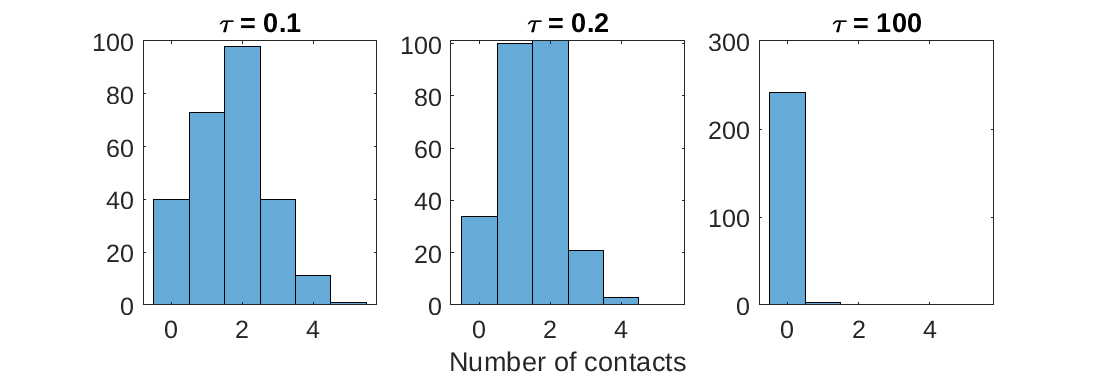}
      \caption{Example distributions of the number of contacts for the volume fractions $f=0.14$ (top row) and $f=0.25$ (bottom row) with different stickiness parameters $\tau$}.
      \label{fig_hist}
\end{figure}

Fig. \ref{fig_hist} shows example distributions of the number of contacts generated by using the placement algorithm for the different stickiness parameters $\tau$ and volume fractions $f$. From the histograms one can see that the mean value of the number of contacts increases not only with the inverse of the stickiness parameter $\tau^{-1}$ but also with the volume fraction $f$. For $\tau = 100$ almost all the particles are unbounded, and therefore, the distribution for large $\tau$ corresponds to that of the random hard sphere deposition without adhesion. 

\subsection{RT-CB code}
The RT-CB code \cite{vaisanen2016} with the SSF correction is available at \\ \url{https://bitbucket.org/planetarysystemresearch/rtcb_public/}. The code is written in Fortran 90 and is parallellized using the Message Parsing Interface (MPI) and Open Multi-Processing (OpenMP) libraries. The code has a simple user interface in which the input parameters, for example, wavelength, medium and monomer sizes, volume fraction, and refractive index are written in the input file. In the new version of the RT-CB code, the input file has a flag to indicate whether or not the SSF correction will be used, and to set the stickiness parameter $\tau$ as follows
\begin{verbatim}
static_structure_factor=.true. (or .false.)
stickiness=0.1
\end{verbatim}

The output files contain the ensemble-averaged Mueller matrix $P$ for the RT and RT-CB solutions, and the scattering $A_{\rm sca}$ and absorption $A_{\rm abs}$ efficiencies. The output Mueller matrix is normalized to $4\pi$ when integrated over the surface of the unit sphere. Since the RT-CB does not account for diffraction, the sum of the efficiencies $A_{\rm sca} + A_{\rm abs} = 1$ . In addition, the SSF-corrected single scattering Mueller matrix, albedo and the mean free path are printed in the output file.

\section{Numerical examples}

In this section, we demonstrate the applicability of the SSF-corrected RT-CB code by comparing the results to the numerically exact fast superposition T-matrix method (FaSTMM) \cite{Markkanen2017} results. Spherical monomers with radii $a$ are randomly deposited into a spherical volume with radius $R$. The number of monomers are chosen so that the target volumetric filling factor $f$ is reached. The medium for the FaSTMM solution is generated using the acceptance-rejection random deposition algorithm in which the spheres are not allowed to overlap. In the SSF-corrected RT-CB solution the medium is approximated by the stickiness parameter $\tau = 1000$ corresponding a non-sticky case. The FaSTMM solution is averaged over 256 random realizations and over 128 scattering planes per realization.

\begin{figure*}[ht]
      \includegraphics[width=0.5\textwidth]{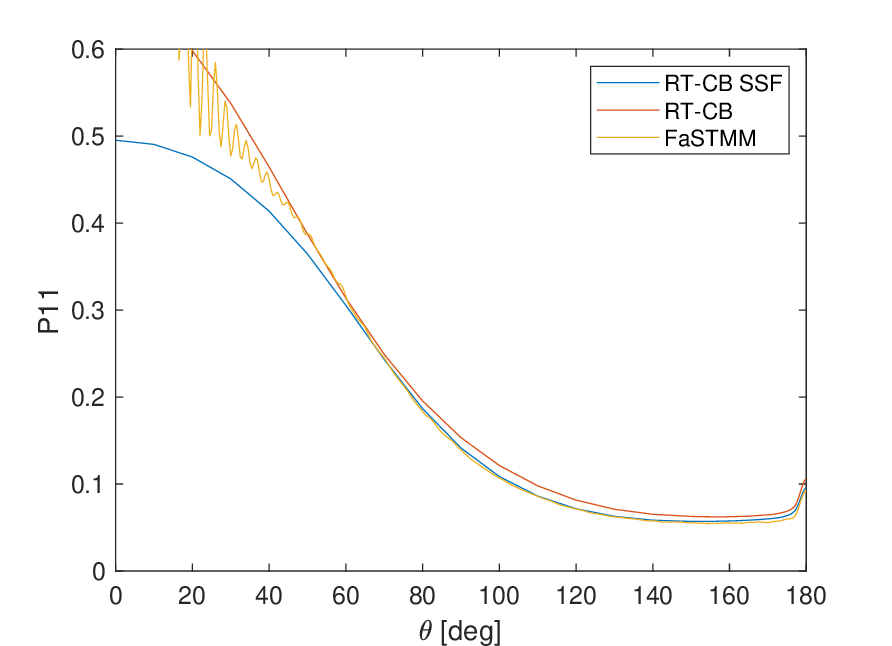}
      \includegraphics[width=0.5\textwidth]{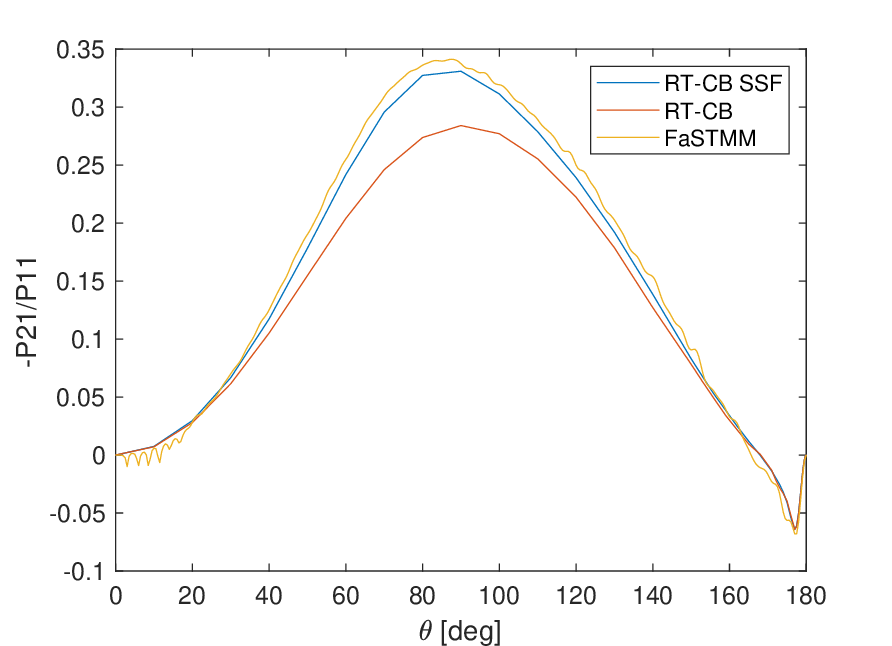}
      \includegraphics[width=0.5\textwidth]{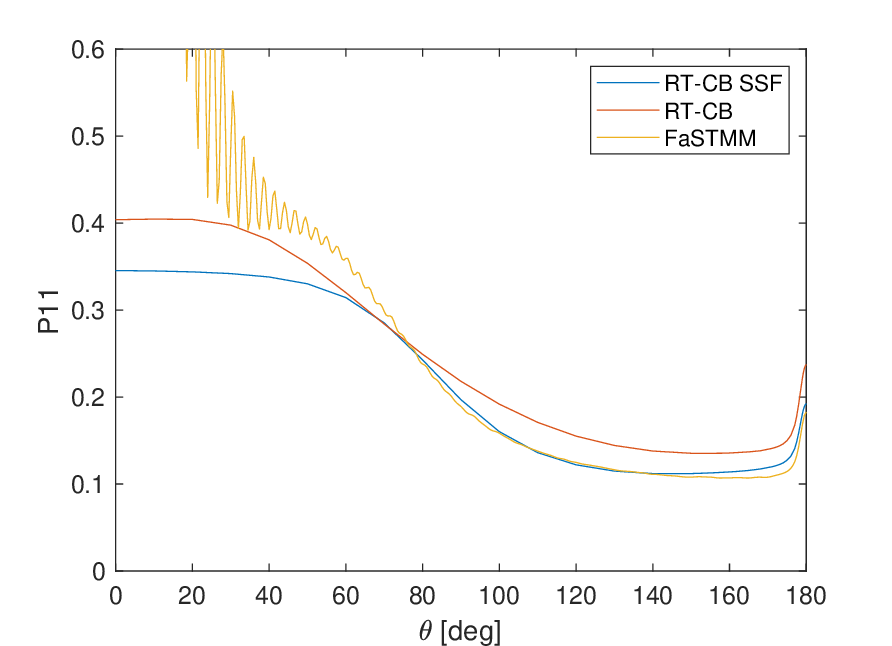}
      \includegraphics[width=0.5\textwidth]{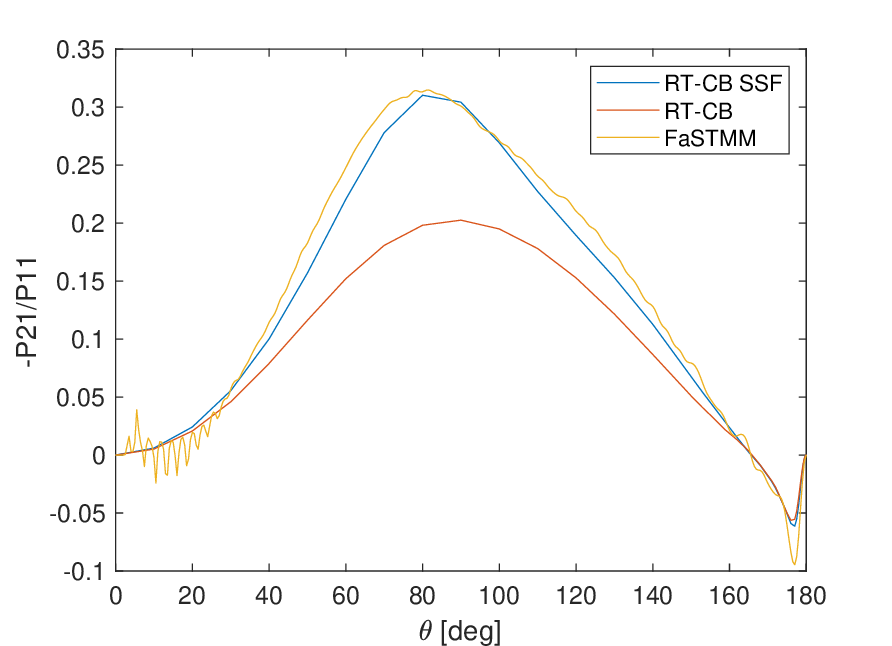}
      \includegraphics[width=0.5\textwidth]{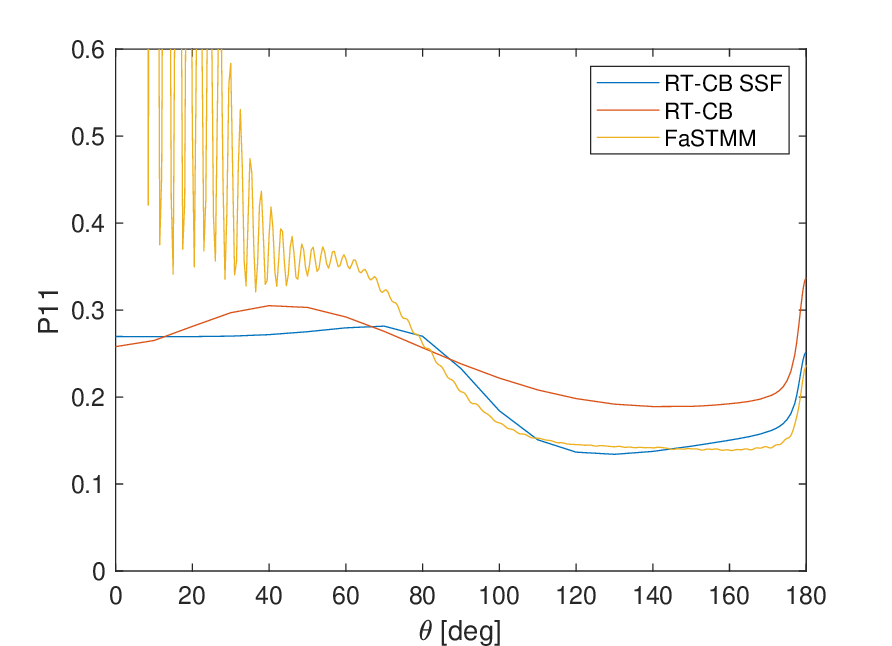}
      \includegraphics[width=0.5\textwidth]{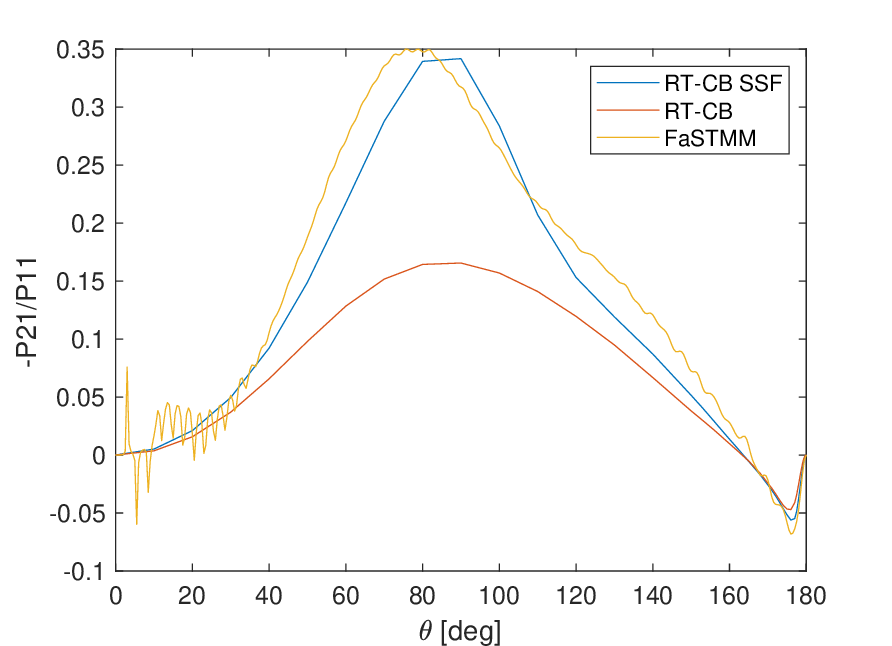}
      \caption{Scattered intensity normalized to the geometric albedo at $\theta = 180^\circ$ (left column), and the degree of linear polarization (right column) computed by using RT-CB with and without the SSF correction for the volume fractions $f=0.1$ (top row), $f=0.2$ (middle row) and  $f=0.3$ (bottom row). Also shown are the numerically exact solutions computed using FaSTMM. The monomer size is \SI{0.15}{\micro\metre} and the wavelength $\lambda =$ \SI{0.45}{\micro\metre}. The refractive index $m = 1.313 + i1.0 \times 10^{-7}$ corresponding water ice.}
      \label{fig1}
\end{figure*}

Fig. \ref{fig1} shows the intensity and the degree of linear polarization as functions of the scattering angle computed by RT-CB and FaSTMM. The computed scattering matrix P11 elements are normalized to the geometric albedo at $\theta = 180^{\circ}$ in order to compare the absolute scattered intensities of the both methods. The output P11 element is normalized to $4\pi$ for RT-CB and $k^2C_{\rm sca}$ for FaSTMM. Hence, the normalization to the geometric albedo is obtained by multiplying the RT-CB scattering matrix by $A_{\rm sca}/4$, and the FaSTMM scattering matrix by $\pi / (k^2 G)$, where $G$ is the geometric cross section. The RT-CB solutions are computed with and without the SSF correction. The wavelength $\lambda = $ \SI{0.45}{\micro\metre} and the monomer size $a =$ \SI{0.15}{\micro\metre}. The refractive index $m = 1.313 + i1.0 \times 10^{-7}$ corresponding water ice. Three different volume filling factors are used $f=0.1$, $f=0.2$ and $f=0.3$. The size of the entire cluster is $R = \SI{5}{\micro\metre}$ corresponding 3700, 7400 and 11100 monomers for $f=0.1$, $f=0.2$ and $f = 0.3$, respectively.

The resulting mean free path lengths are $\ell^{0.1}_{\rm Mie} = \SI{2.762}{\micro\metre}$, $\ell^{0.2}_{\rm Mie} = \SI{1.381}{\micro\metre}$ and $\ell^{0.3}_{\rm Mie} = \SI{0.921}{\micro\metre}$ without, and $\ell^{0.1}_{\rm SSF} = \SI{3,875}{\micro\metre}$,  $\ell^{0.2}_{\rm SSF} = \SI{2.827}{\micro\metre}$ and $\ell^{0.3}_{\rm SSF} = \SI{2.920}{\micro\metre}$ with the SSF correction for $f=0.1$, 0.2 and 0.3, respectively.
It is evident that the SSF correction improves the solution accuracy. The solutions without the SSF correction overestimate the intensity and underestimate the degree of polarization. Such behavior is understandable as the SSF correction increases the mean free path length, and thus, decreases intensity and increases polarization. Also, we find that the RT-CB code can reproduce the coherent backscattering effect, i.e., the intensity spike and the negative polarization branch near the backscattering direction ($\theta > 160^{\circ}$) as expected. Although, the solution underestimates the minimum value of the negative polarization for the denser configurations, and the effect of the SSF correction around the backscattering direction is rather small. This discrepancy may arise  because the SSF correction does not take the near-field effects into account. The RT-CB code cannot reproduce the intensity around the forward scattering direction because the coherent part of the scattered radiation giving rise to diffraction is not taken into account. The other elements of the Mueller matrix also show similar agreement.

\begin{figure*}[h]
      \includegraphics[width=0.5\textwidth]{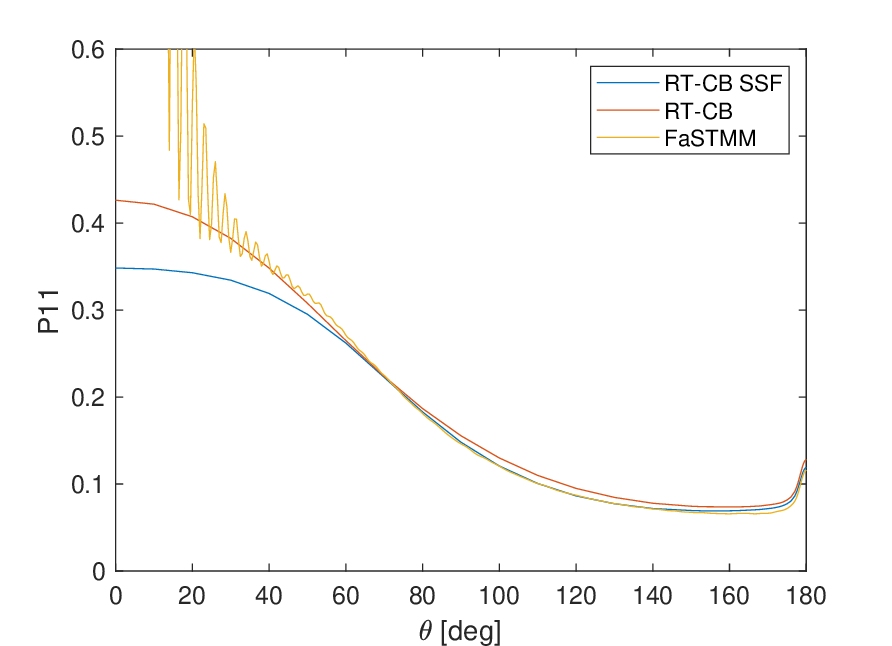}
      \includegraphics[width=0.5\textwidth]{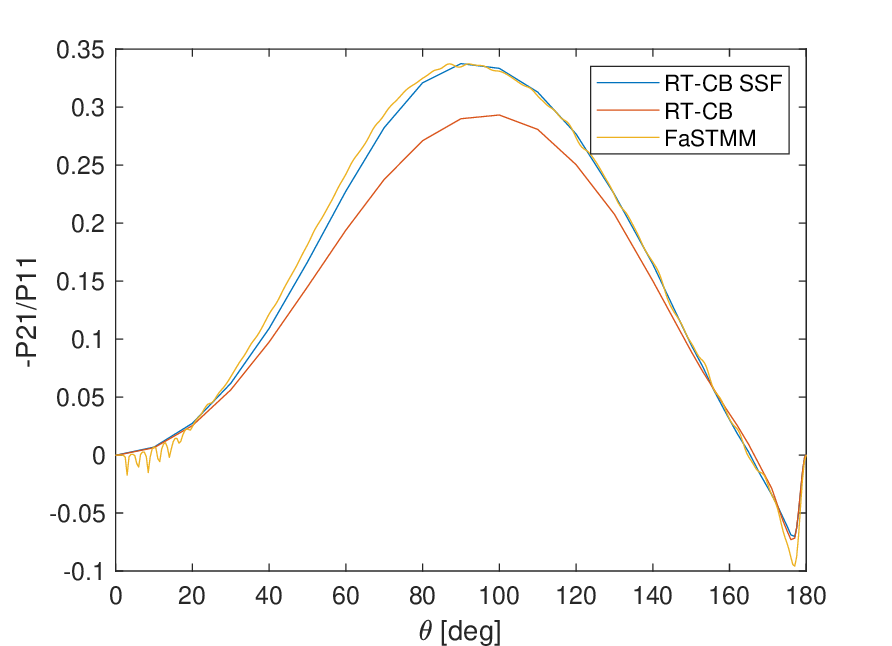}
      \caption{Same as in Fig. \ref{fig1} but for a core-mantle monomer structure with $f = 0.1$.}
      \label{fig3}
\end{figure*}

Next, we investigate monomers having a core-mantle structure. We set the core radius $a_{\rm c} = a/2$ with $a = \SI{0.15}{\micro\metre}$, and the refractive index of the core $m_{\rm c} = 1.6 + i0.05$ and the mantle $m = 1.313 + i1.0 \times 10^{-7}$. The mean free path and the single scattering albedo without the SSF correction are $\omega_{\rm Mie} = 0.9417$ and $\ell_{\rm Mie} = \SI{1.958}{\micro\metre}$, and with the SSF correction $\omega_{\rm SSF} = 0.9216$ and $\ell_{\rm SSF} = \SI{2.635}{\micro\metre}$.  In Fig. \ref{fig3}, we observe good agreement between the SSF-corrected RT-CB method and the numerically exact FaSTMM method. The core-mantle monomer structure is also a new feature in the extended RT-CB code. It may be useful for some  astrophysical applications as interstellar dust grains are believed to consist of monomers with silicate core and organic or icy mantles \cite{greenberg}.

\begin{figure*}[htb]
      \includegraphics[width=0.5\textwidth]{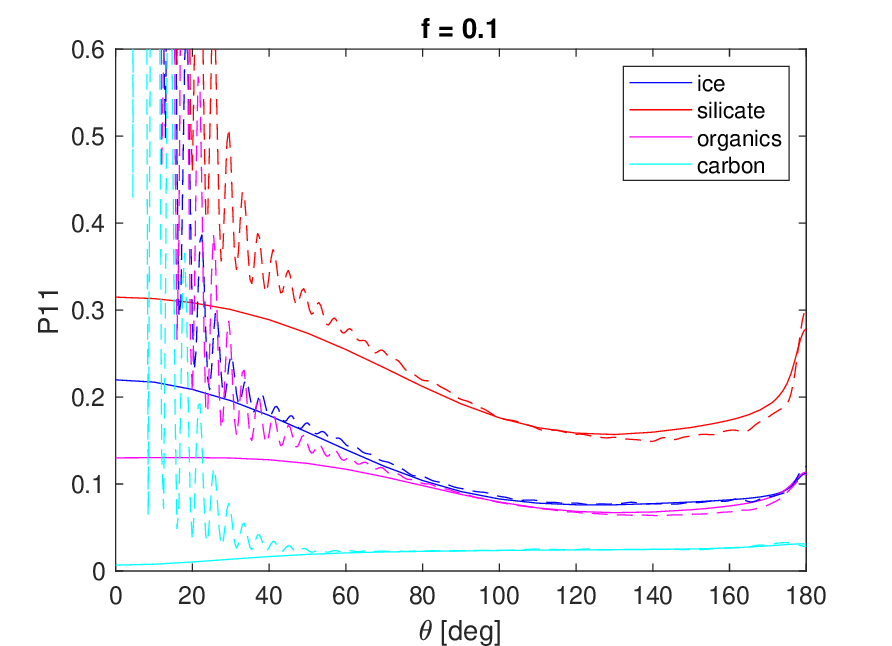}
      \includegraphics[width=0.5\textwidth]{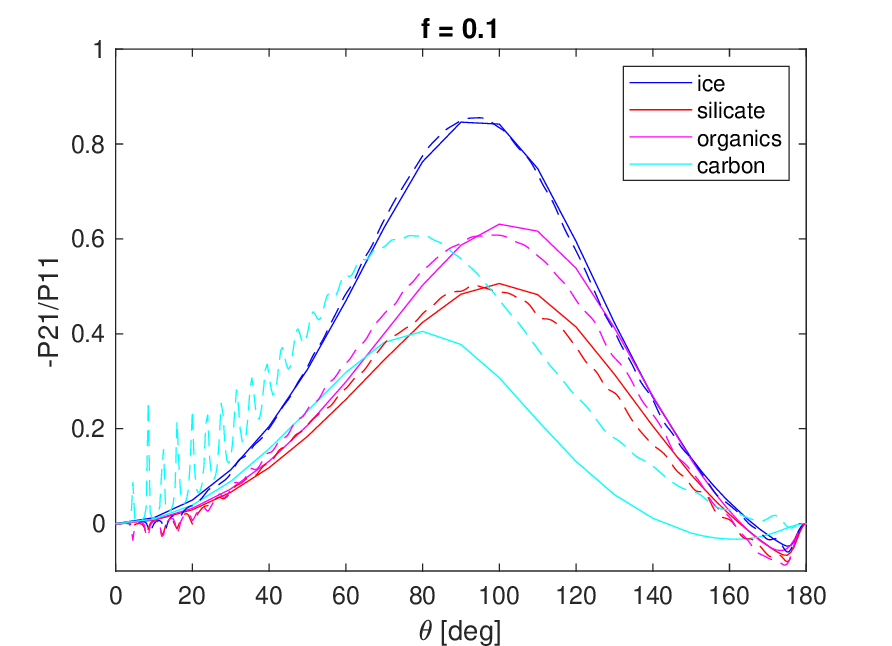}
       \includegraphics[width=0.5\textwidth]{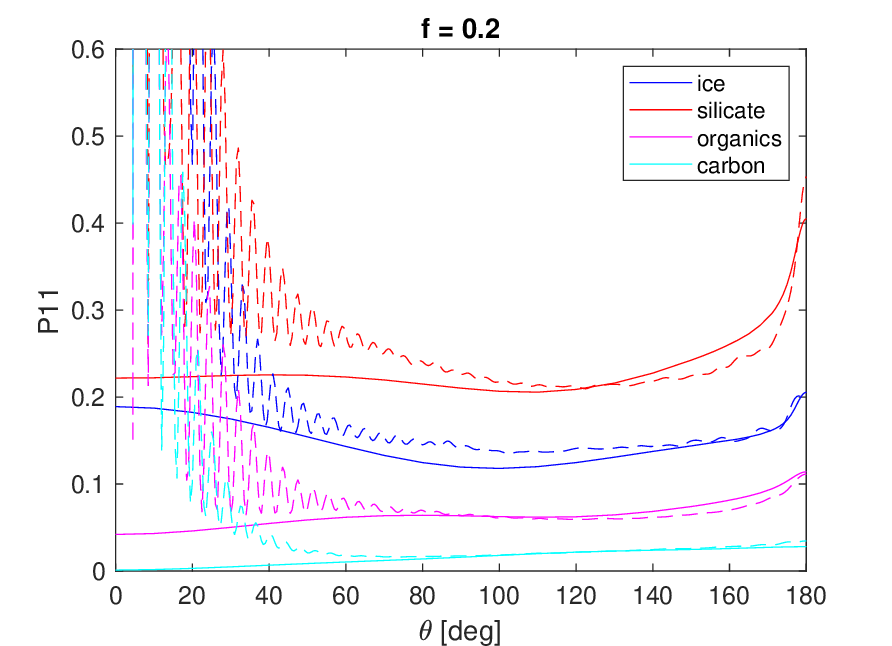}
      \includegraphics[width=0.5\textwidth]{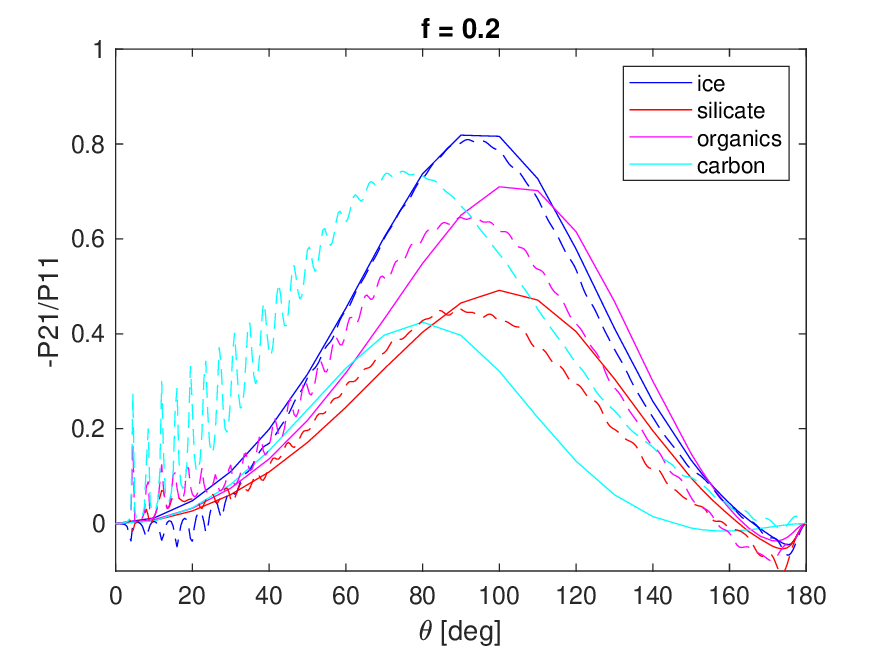}
      \caption{Scattered intensity (left column) and the degree of linear polarization (right column) for different materials computed by SSF-corrected RT-CB (solid lines) and FaSTMM (dashed lines). The wavelength $\lambda = \SI{0.65}{\micro\metre}$, the monomer size $a = \SI{0.15}{\micro\metre}$ and the medium radius $R=\SI{5}{\micro\metre}$.}
      \label{fig_mat}
\end{figure*}

Next, we demonstrate that the solution accuracy of the SSF-corrected RT-CB deteriorates with increasing refractive index. Fig. \ref{fig_mat} shows the intensity and the degree of linear polarization for different refractive indices. The refractive indices are chosen to correspond typical material compositions of dust in the solar system in visible light: water ice with $m = 1.313 + i1\times 10^{-7}$, silicate mineral with $m = 1.6 + i0.0001$, organic refractory with $m = 1.6 + i0.05$, and amorphous carbon with $m  = 2.0 + i0.8$. The wavelength $\lambda = \SI{0.65}{\micro\metre}$, the monomer radius $a=\SI{0.15}{\micro\metre}$, and the medium radius $R=\SI{5}{\micro\metre}$. The solid lines indicate the SSF-corrected RT-CB solutions and the dashed lines indicate the FaSTMM solutions.

It is clear from Fig. \ref{fig_mat} that for high-contrast material such as amorphous carbon the accuracy of the SSF-corrected RT-CB is not good. In particular, the degree of linear polarization shows a large difference compared to the numerically exact FaSTMM solution. This is expected as for the high-contrast material the mean-free-path length is comparable to wavelength as indicated in Table \ref{table1}. Hence, the near-field effects becomes more dominant which are not accounted for in the SSF-corrected RT-CB solution. The SSF-corrected RT-CB computes the ladder and cyclical diagrams modified with the pair distribution function. The modification only accounts for the far-field interferences arising from the correlated positions of scatterers. Further, interactions in the scattering sequences are calculated using the far-field approximation. In addition, the boundary effects arising from the coherent field are neglected which also become more important with higher refractive index and volume fractions. 

\begin{table}
  \center
  \caption{SSF-corrected single scattering albedos and mean free paths lengths for $f=0.1$ and 0.2}
\begin{tabular}[HT]{|l|c|c|c|c|c}
\hline
  & $\omega^{0.1}_{\rm SSF}$ & $\ell^{0.1}_{\rm SSF}$ & $\omega^{0.2}_{\rm SSF}$ & $\ell^{0.2}_{\rm SSF}$ \\
  \hline
  ice  & 1.0 & \SI{10.83}{\micro\metre} & 1.0 & \SI{8.04}{\micro\metre}\\
  \hline
  silicate  & 0.991 & \SI{2.973}{\micro\metre} & 0.986 & \SI{2.35}{\micro\metre}\\
  \hline
  organics  & 0.695 & \SI{2.260}{\micro\metre} & 0.587 & \SI{1.53}{\micro\metre}\\
  \hline
  carbon  & 0.355 & \SI{0.798}{\micro\metre} & 0.264 & \SI{0.46}{\micro\metre} \\
  \hline
  
\end{tabular}
\label{table1}
\end{table}

\begin{figure*}[htb]
      \includegraphics[width=0.5\textwidth]{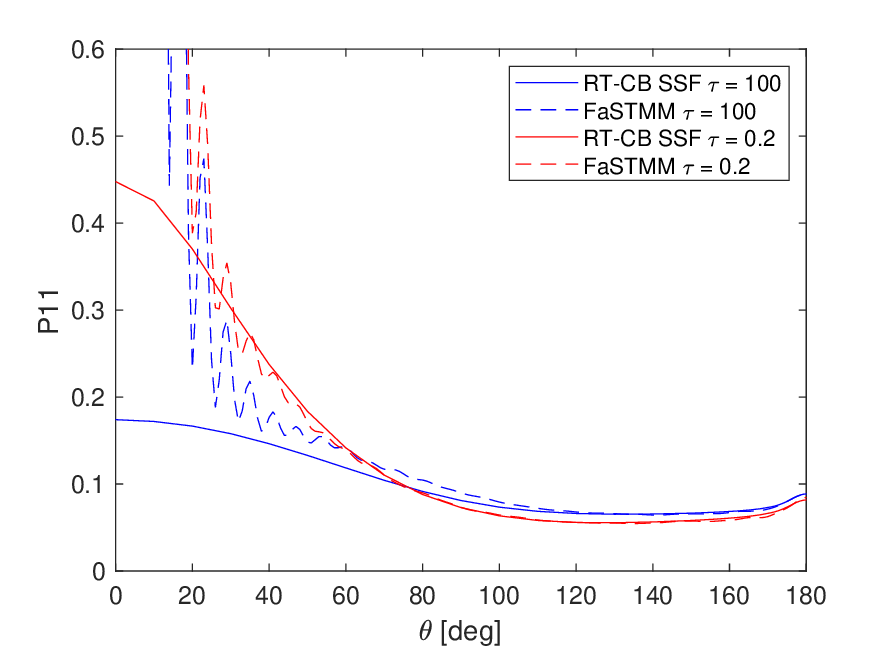}
      \includegraphics[width=0.5\textwidth]{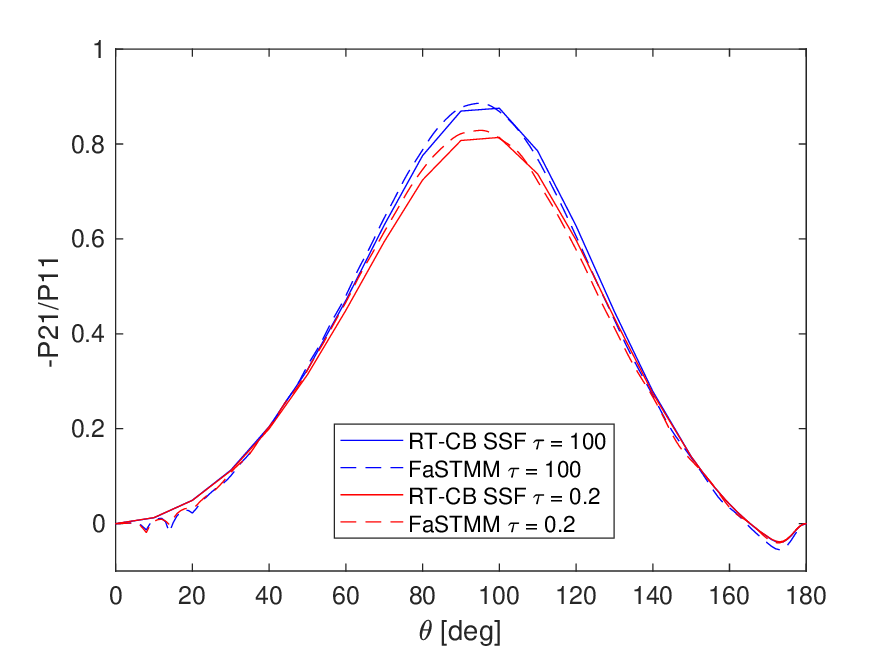}
      \caption{Effect of the stickiness parameter $\tau$ on the intensity and the degree of linear polarization for $f = 0.14$. $\tau=100$ corresponds non-sticky case, and $\tau=0.2$ moderate stickiness. The radius of the medium is $R=\SI{10}{\micro\metre}$, radius of spheres $a=\SI{0.5}{\micro\metre}$ and $\lambda = \SI{2.0944}{\micro\metre}$.}
      \label{fig4}
\end{figure*}

Finally, we demonstrate the effects of the stickiness parameter $\tau$ in Fig. \ref{fig4}. The medium $R= \SI{10}{\micro\metre}$ consists of monomers of radius $a = \SI{0.5}{\micro\metre}$ and $\lambda = \SI{2.0944}{\micro\metre}$. The refractive index is $m = 1.313 + i1.0 \times 10^{-7}$. Two different stickiness parameters are considered $\tau = 100$ and 0.2. When the stickiness parameter is large, $\tau = 100$, the surface adhesion is negligible and the monomer arrangement corresponds to that of random packing of hard spheres. With decreasing $\tau$ the attractive force among monomers increases, and the spheres form aggregates where some of the monomers are connected. The smallest $\tau$ that can be used with the Percus-Yevick approximation is around $\tau = 0.1$. The pair distribution functions and the structure factors for these cases are shown in Fig. \ref{fig0}. The resulting mean free path lengths for $\tau=100$ and 0.2 are $\ell_{100}=\SI{32.99}{\micro\metre}$ and  $\ell_{0.2}=\SI{26.46}{\micro\metre}$, respectively. The results for SSF-corrected RT-CB are plotted as solid lines and for FaSTMM as dashed lines. The algorithm presented in Section \ref{sec_placement} was used to generate the medium for the FaSTMM computations. The RT-CB and FaSTMM results show almost a perfect match. For smaller $\tau$ the packing algorithm became too slow and we could not generate the reference result with FaSTMM.
Fig. \ref{fig4} shows that decreasing $\tau$ decreases intensity in the backscattering direction but increases intensity in the forward direction. This is evident as the static structure factor $S_{\rm M}$ shows increasing forward scattering with decreasing $\tau$ as demonstrated in Fig. \ref{fig0}. The mean free path length decreases with decreasing $\tau$ which explains decreasing polarization with decreasing $\tau$.

\section{Conclusions}

We have presented an extended version of the RT-CB code where the correlated positions of monomers are accounted for using the sticky-hard-sphere Percus-Yevick approximation. The code is available as open source in \url{https://bitbucket.org/planetarysystemresearch/rtcb_public/}. The new version improves the accuracy of  solution when the packing density is moderate or high. In particular, we demonstrated that the solution compares well with the exact numerical solution when the refractive index is close to that of the background medium, for example icy particles in visible wavelengths. For a higher refractive index or packing density, we demonstrated that the solution accuracy deteriorates as the SSF correction does not account for the near-zone nor boundary effects. Thus, we strongly recommend comparing the RT-CB solution to the numerically exact solution before applying it to a new type of problems to estimate potential errors. In general, good solution accuracy is expected if the mean-free-path length is much longer than the wavelength.

In addition, the new version allows for treating media where monomers are adhesive and can be in contact with each other. The adhesion is implemented by using the sticky-hard-sphere Percus-Yevick approximation where the strength of adhesive force is controlled by the stickiness parameter. The RT-CB solution for adhesive icy spheres shows a good agreement with the numerically exact FaSTMM solution. 


\section*{Acknowledgements}
This work was supported by the ERC Starting Grant No. 757390 and the German Research Foundation DFG Grant No. 517146316. Computational resources have been provided by Gesellschaft f\"ur Wissenschaftliche Datenverarbeitung mbH G\"ottingen (GWDG).


 \bibliographystyle{elsarticle-num} 
 \bibliography{cas-refs}

\begin{thebibliography}{10}
\expandafter\ifx\csname url\endcsname\relax
  \def\url#1{\texttt{#1}}\fi
\expandafter\ifx\csname urlprefix\endcsname\relax\def\urlprefix{URL }\fi
\expandafter\ifx\csname href\endcsname\relax
  \def\href#1#2{#2} \def\path#1{#1}\fi

\bibitem{penttila2021}
A.~Penttilä, J.~Markkanen, T.~Väisänen, J.~Räbinä, M.~A. Yurkin,
  K.~Muinonen,
 {How
  much is enough? the convergence of finite sample scattering properties to
  those of infinite media}, Journal of Quantitative Spectroscopy and Radiative
  Transfer 262 (2021) 107524.
\newblock \href {https://doi.org/https://doi.org/10.1016/j.jqsrt.2021.107524}
  {\path{doi:https://doi.org/10.1016/j.jqsrt.2021.107524}}.


\bibitem{MISHCHENKO2016}
M.~I. Mishchenko, J.~M. Dlugach, M.~A. Yurkin, L.~Bi, B.~Cairns, L.~Liu, R.~L.
  Panetta, L.~D. Travis, P.~Yang, N.~T. Zakharova,
 {First-principles
  modeling of electromagnetic scattering by discrete and discretely
  heterogeneous random media}, Physics Reports 632 (2016) 1--75,
  first-principles modeling of electromagnetic scattering by discrete and
  discretely heterogeneous random media.
\newblock \href {https://doi.org/https://doi.org/10.1016/j.physrep.2016.04.002}
  {\path{doi:https://doi.org/10.1016/j.physrep.2016.04.002}}.


\bibitem{DOICU2019}
A.~Doicu, M.~I. Mishchenko,
 {An
  overview of methods for deriving the radiative transfer theory from the
  maxwell equations. ii: Approach based on the dyson and bethe–salpeter
  equations}, Journal of Quantitative Spectroscopy and Radiative Transfer 224
  (2019) 25--36.
\newblock \href {https://doi.org/https://doi.org/10.1016/j.jqsrt.2018.10.032}
  {\path{doi:https://doi.org/10.1016/j.jqsrt.2018.10.032}}.


\bibitem{TISHKOVETS2003}
V.~P. Tishkovets, M.~I. Mishchenko,
 {Coherent
  backscattering of light by a layer of discrete random medium}, Journal of
  Quantitative Spectroscopy and Radiative Transfer 86~(2) (2004) 161--180.
\newblock \href {https://doi.org/https://doi.org/10.1016/S0022-4073(03)00281-4}
  {\path{doi:https://doi.org/10.1016/S0022-4073(03)00281-4}}.


\bibitem{Muinonen2004}
  K.~Muinonen,
  {Coherent
  backscattering of light by complex random media of spherical scatterers:
  numerical solution}, Waves in Random Media 14~(3) (2004) 365--388.
\newblock \href
  {https://doi.org/10.1088/0959-7174/14/3/010}
  {\path{doi:10.1088/0959-7174/14/3/010}}.


\bibitem{TISHKOVETS2020}
V.~P. Tishkovets, E.~V. Petrova,
  {An
  algorithm and codes for fast computations of the opposition effects in a
  semi-infinite discrete random medium}, Journal of Quantitative Spectroscopy
  and Radiative Transfer 255 (2020) 107252.
\newblock \href {https://doi.org/https://doi.org/10.1016/j.jqsrt.2020.107252}
  {\path{doi:https://doi.org/10.1016/j.jqsrt.2020.107252}}.


\bibitem{Muinonen2012}
K.~Muinonen, M.~I. Mishchenko, J.~M. Dlugach, E.~Zubko, A.~Penttilä,
G.~Videen,
{COHERENT
  BACKSCATTERING VERIFIED NUMERICALLY FOR a FINITE VOLUME OF
  SPHERICAL PARTICLES}, The Astrophysical Journal 760 (2) (2012) 118.
\newblock \href {https://doi.org/10.1088/0004-637x/760/2/118}
  {\path{doi:10.1088/0004-637x/760/2/118}}.


\bibitem{vaisanen2016}
T.~Väisänen, A.~Penttilä, J.~Markkanen, K.~Muinonen, Validation of radiative
  transfer and coherent backscattering for discrete random media, in: 2016 URSI
  International Symposium on Electromagnetic Theory (EMTS), 2016, pp. 396--399.
\newblock \href {https://doi.org/10.1109/URSI-EMTS.2016.7571408}
  {\path{doi:10.1109/URSI-EMTS.2016.7571408}}.

\bibitem{Cartigny1986}
J.~D. Cartigny, Y.~Yamada, C.~L. Tien,
 {Radiative Transfer With Dependent
  Scattering by Particles: Part 1 - Theoretical Investigation}, Journal of
  Heat Transfer 108~(3) (1986) 608--613.
\newblock 
  \href {https://doi.org/10.1115/1.3246979} {\path{doi:10.1115/1.3246979}}.


\bibitem{Drolen1987}
B.~L. Drolen, C.~L. Tien, {Independent and
  dependent scattering in packed-sphere systems}, Journal of Thermophysics and
  Heat Transfer 1~(1) (1987) 63--68.
\newblock \href
  {https://doi.org/10.2514/3.8} {\path{doi:10.2514/3.8}}.

\bibitem{Tsang1985}
L.~Tsang, J.~A. Kong, R.~T. Shin, Theory of microwave remote sensing (1985).

\bibitem{MISHCHENKO1994}
M.~I. Mishchenko,
 {Asymmetry
  parameters of the phase function for densely packed scattering grains},
  Journal of Quantitative Spectroscopy and Radiative Transfer 52~(1) (1994)
  95--110.
\newblock \href {https://doi.org/https://doi.org/10.1016/0022-4073(94)90142-2}
  {\path{doi:https://doi.org/10.1016/0022-4073(94)90142-2}}.


\bibitem{Tsang2000}
L.~Tsang, C.-T. Chen, A.~T.~C. Chang, J.~Guo, K.-H. Ding, Dense media radiative
  transfer theory based on quasicrystalline approximation with applications to
  passive microwave remote sensing of snow, Radio Science 35~(3) (2000)
  731--749.
\newblock \href {https://doi.org/10.1029/1999RS002270}
  {\path{doi:10.1029/1999RS002270}}.

\bibitem{TISHKOVETS2006}
V.~P. Tishkovets, K.~Jockers,
 {Multiple
  scattering of light by densely packed random media of spherical particles:
  Dense media vector radiative transfer equation}, Journal of Quantitative
  Spectroscopy and Radiative Transfer 101~(1) (2006) 54--72.
\newblock \href {https://doi.org/https://doi.org/10.1016/j.jqsrt.2005.10.001}
  {\path{doi:https://doi.org/10.1016/j.jqsrt.2005.10.001}}.


\bibitem{Tsang2007}
L.~Tsang, J.~Pan, D.~Liang, Z.~Li, D.~W. Cline, Y.~Tan, Modeling active
  microwave remote sensing of snow using dense media radiative transfer (dmrt)
  theory with multiple-scattering effects, IEEE Transactions on Geoscience and
  Remote Sensing 45~(4) (2007) 990--1004.
\newblock \href {https://doi.org/10.1109/TGRS.2006.888854}
  {\path{doi:10.1109/TGRS.2006.888854}}.

\bibitem{ito2018}
G.~Ito, M.~I. Mishchenko, T.~D. Glotch, Radiative-transfer modeling of spectra
  of planetary regoliths using cluster-based dense packing modifications,
  Journal of Geophysical Research: Planets 123~(5) (2018) 1203--1220.

\bibitem{Ma2020}
L.~Ma, C.~Wang, L.~Liu,
 {Polarized
  radiative transfer in dense dispersed media containing optically soft sticky
  particles}, Opt. Express 28~(19) (2020) 28252--28268.
\newblock \href {https://doi.org/10.1364/OE.404121}
  {\path{doi:10.1364/OE.404121}}.


\bibitem{Zurk1996}
L.~M. Zurk, L.~Tsang, D.~P. Winebrenner,
 {Scattering properties of dense
  media from monte carlo simulations with application to active remote sensing
  of snow}, Radio Science 31~(4) (1996) 803--819.
\newblock \href {https://doi.org/10.1029/96RS00939}
  {\path{doi:10.1029/96RS00939}}.

\bibitem{TISHKOVETS2013}
V.~P. Tishkovets, E.~V. Petrova,
  {Coherent
  backscattering by discrete random media composed of clusters of spherical
  particles}, Journal of Quantitative Spectroscopy and Radiative Transfer 127
  (2013) 192--206.
\newblock \href {https://doi.org/https://doi.org/10.1016/j.jqsrt.2013.05.017}
  {\path{doi:https://doi.org/10.1016/j.jqsrt.2013.05.017}}.


\bibitem{Markkanen2018_2}
J.~Markkanen, J.~Agarwal, T.~Väisänen, A.~Penttilä, K.~Muinonen,
 {Interpretation of the
  phase functions measured by the {OSIRIS} instrument for comet
  67p/churyumov{\textendash}gerasimenko}, APJL 868~(1) (2018) L16.
\newblock \href {https://doi.org/10.3847/2041-8213/aaee10}
  {\path{doi:10.3847/2041-8213/aaee10}}.

\bibitem{Muinonen2018}
K.~Muinonen, J.~Markkanen, T.~V\"{a}is\"{a}nen, J.~Peltoniemi, A.~Penttil\"{a},
 {Multiple scattering of
  light in discrete random media using incoherent interactions}, Opt. Lett.
  43~(4) (2018) 683--686.
\newblock \href {https://doi.org/10.1364/OL.43.000683}
  {\path{doi:10.1364/OL.43.000683}}.

\bibitem{Markkanen2018}
J.~Markkanen, T.~V\"{a}is\"{a}nen, A.~Penttil\"{a}, K.~Muinonen,
 {Scattering and
  absorption in dense discrete random media of irregular particles}, Opt. Lett.
  43~(12) (2018) 2925--2928.
\newblock \href {https://doi.org/10.1364/OL.43.002925}
  {\path{doi:10.1364/OL.43.002925}}.

\bibitem{Vaisanen2019}
T.~Väisänen, J.~Markkanen, A.~Penttilä, K.~Muinonen,
 {Radiative transfer with
  reciprocal transactions: Numerical method and its implementation}, PLOS ONE
  14~(1) (2019) 1--24.
\newblock \href {https://doi.org/10.1371/journal.pone.0210155}
  {\path{doi:10.1371/journal.pone.0210155}}.


\bibitem{Baxter1968}
R.~J. Baxter, {Percus–yevick equation
  for hard spheres with surface adhesion}, The Journal of Chemical Physics
  49~(6) (1968) 2770--2774.
\newblock \href {http://arxiv.org/abs/https://doi.org/10.1063/1.1670482}
  {\path{arXiv:https://doi.org/10.1063/1.1670482}}, \href
  {https://doi.org/10.1063/1.1670482} {\path{doi:10.1063/1.1670482}}.


\bibitem{Tsang2001}
L.~Tsang, J.~Kong, K.~Ding, O.~Ao, Scattering of Electromagnetic Waves,
  Numerical Simulations, Vol.~2, 2001.
\newblock \href {https://doi.org/10.1002/0471224308}
  {\path{doi:10.1002/0471224308}}.

\bibitem{Wertheim}
M.~S. Wertheim,
 {Exact solution of
  the percus-yevick integral equation for hard spheres}, Phys. Rev. Lett. 10
  (1963) 321--323.
\newblock \href {https://doi.org/10.1103/PhysRevLett.10.321}
  {\path{doi:10.1103/PhysRevLett.10.321}}.


\bibitem{Hovenier94}
J.~W. Hovenier,
{Structure of
  a general pure mueller matrix}, Appl. Opt. 33~(36) (1994) 8318--8324.
\newblock \href {https://doi.org/10.1364/AO.33.008318}
  {\path{doi:10.1364/AO.33.008318}}.


\bibitem{Seaton1987}
N.~A. Seaton, E.~D. Glandt, {Monte
  Carlo simulation of adhesive spheres}, The Journal of Chemical Physics
  87~(3) (1987) 1785--1790.
\newblock
  \href {https://doi.org/10.1063/1.453724} {\path{doi:10.1063/1.453724}}.


\bibitem{Kranendonk1988}
W.~Kranendonk, D.~Frenkel,
 {Simulation of the
  adhesive-hard-sphere model}, Molecular Physics 64~(3) (1988) 403--424.
 \newblock
 \href
  {https://doi.org/10.1080/00268978800100303}
  {\path{doi:10.1080/00268978800100303}}.

\bibitem{Markkanen2017}
J.~Markkanen, A.~J. Yuffa,
 {Fast
  superposition t-matrix solution for clusters with arbitrarily-shaped
  constituent particles}, Journal of Quantitative Spectroscopy and Radiative
  Transfer 189 (2017) 181 -- 188.
\newblock \href {https://doi.org/https://doi.org/10.1016/j.jqsrt.2016.11.004}
  {\path{doi:https://doi.org/10.1016/j.jqsrt.2016.11.004}}.


\bibitem{greenberg}
J.~{Greenberg}, {The Core-Mantle Model of Interstellar Grains and the Cosmic
  Dust Connection}, in: L.~J. {Allamandola}, A.~G.~G.~M. {Tielens} (Eds.),
  Interstellar Dust, Vol. 135, 1989, p. 345.

\end{thebibliography}





\end{document}